\documentclass{PoS}

\usepackage{amsmath}
\usepackage{amsfonts}
\usepackage{amssymb}
\usepackage{graphicx}
\usepackage{fontenc}
\usepackage{times}
\usepackage{mathptmx}

\newcommand{\Tr}{\mathrm{Tr}}

\newcommand{\comma}{\; ,}
\newcommand{\beq}{\begin{eqnarray}}
\newcommand{\eeq}{\end{eqnarray}}

\newcommand{\VEV}[1]{\left\langle #1 \right\rangle}
\newcommand{\RC}{\mathcal R}
\newcommand{\hPh}{\hat{\Phi}}

\title{The glueball spectrum at large N}

\ShortTitle{The glueball spectrum at large N}

\author{Biagio Lucini \\
  School of Physical Sciences, Swansea University\\
  Singleton Park, Swansea SA2 8PP, UK\\
  E-mail: \email{b.lucini@swansea.ac.uk} 
}
\author{Antonio Rago\\
  Department of Physics, Bergische Universit\"at Wuppertal\\
  Gaussstr. 20, D-42119 Wuppertal, Germany\\
  E-mail: \email{rago@physik.uni-wuppertal.de} 
}
\author{\speaker{Enrico Rinaldi}\\
  SUPA, School of Physics and Astronomy, University of Edinburgh\\
  Edinburgh EH9 3JZ, UK\\
  E-mail: \email{e.rinaldi@sms.ed.ac.uk} 
}

\abstract{ The lowest-lying glueball masses are computed in SU($N$) gauge theory 
  on a spacetime lattice for constant value of the lattice spacing $a$
  and for $N$ ranging from $3$ to $8$. The lattice spacing is fixed
  using the deconfinement temperature at temporal extension of the
  lattice $N_T = 6$. The calculation is conducted employing in each
  channel a variational ansatz performed on a large basis of operators
  that includes also torelon and (for the lightest states) scattering
  trial functions. This basis is constructed using an automatic algorithm that
  allows us to build operators of any size and shape in any irreducible
  representation of the cubic group. A good signal is extracted for
  the ground state and the first excitation in several symmetry
  channels. It is shown that all the observed states are well
  described by their large $N$ values, with modest ${\cal O}(1/N^2)$
  corrections. In addition spurious states are identified that couple
  to torelon and scattering operators.\\
  [0.5cm]
  \rightline{WUB/10-28}
}

\FullConference{The XXVIII International Symposium on Lattice Field
  Theory, Lattice2010\\
		June 14-19, 2010\\
		Villasimius, Italy}

\begin{document}
\section{Introduction}
\label{sect:introduction}
SU($N$) gauge theories in the large $N$ limit
play a central role in the gauge-gravity correspondence and have
become the subject of a line of numerical investigations on the
lattice. In addition to determining values for observables in the
large $N$ limit, lattice calculations provide their corrections at
finite $N$. As dictated by the diagrammatic
expansion~\cite{Hooft:1973jz}, these corrections can be expressed as a
power series in $1/N^2$ for
the quenched theory and in $1/N$ in the dynamical case. The emerging
picture is that at least for the quenched theory only the leading
correction of ${\cal O}(1/N^2)$ is sufficient to describe the system
at any finite value of $N$ bigger than two at a level of accuracy of
the order of a few percents. In order to assess the reliability of various analytical
methods based on the large $N$ framework (which often have to resort
to other approximations in addition to taking the large $N$ limit), it
is important to compare their predictions to the lattice data for
observables that are well under control in both approaches. The
glueball spectrum in the pure Yang-Mills theory is one of the easiest
observables to compare. Previous numerical calculations at large $N$ have been
recently reviewed in~\cite{Teper:2008yi}.\\
In this work we provide the first determination of the large $N$
glueball spectrum (obtained with an extrapolation including values of
$N$ up to eight) in several irreducible representations of the lattice
rotational group and for both values of parity and charge
conjugation. We are also able to
disentangle genuine single-particle states from spurious or
multi-particle resonances that are present on finite volume lattice
simulations.

\section{The method}
\label{sec:method}

The lattice discretisation of SU($N$) Yang-Mills theory used
throughout this work is entirely conventional. We consider the system
defined on an isotropic four-dimensional torus of linear size $L$. If
$a$ is the lattice spacing, the number of points in each direction is
given by $N_L = L/a$. We used the Wilson action for the lattice
theory, given by
\begin{equation}
S = \beta \sum_{i,\mu > \nu} \left( 1 - \frac{1}{N} \mbox{Re} \ \Tr
  \left( U_{\mu \nu}(i) \right) \right) \comma
\end{equation}
where $U_{\mu \nu}(i)$ is the parallel transport of the link variables
along the elementary lattice plaquette and $\beta$ is defined as
$\beta = 2N/g_0^2$, with $g_0$ the bare gauge coupling. In order to
compare quantities at fixed lattice spacing across
different SU($N$) groups, it proves useful to set the scale using the
(pseudo--)critical coupling of the deconfinement transition at fixed
temporal extent $N_T = 6$. A $N_L = 12$ lattice for $\beta =
\beta_c(N_T = 6)$ gives a glueball spectrum in the scaling region and
free from large finite size artefacts~\cite{Lucini:2004my}.\\
In general, masses of bound states on the lattice are extracted from
the exponential decay of connected correlation functions between
operators with the desired quantum numbers. In lattice Yang-Mills theory
these operators are constructed using traces of path ordered products
of links around closed loops. Moreover, the links used in the
operators are smeared and blocked~\cite{Albanese:1987ds,Teper:1987wt}
several times in order to obtain smooth operators on physical length
scales that project onto the low--lying states of the spectrum.
A variational ansatz for the correlators is also employed: for every
set of quantum numbers $J^{PC}$ we measure a matrix of correlators between
different operators and we look for their
linear combination that has the best overlap onto the state we are
interested in. This allows
us to obtain the mass of the groundstate and of the first excitations
of the spectrum
with the smallest possible systematic errors (for more details on the
variational technique see Ref.~\cite{Lucini:2010nv}).\\
At finite volume, the single--particle glueball spectrum receives
non--negligible corrections from multi--glueball states. Moreover, when the
system is closed with periodic boundary conditions (like in our case)
topological excitations wrapping the compact direction ({\em torelons}) with
the same quantum numbers of glueballs appear; if not correctly accounted for,
these states can affect significantly the measured glueball spectrum. 
In order to control these spurious contributions, we include in the
variational set operators that best overlap with two--glueball and torelon
states.

\section{The operators}
\label{sec:operators}

\begin{figure}[ht!]
 \begin{minipage}[0.15\textwidth]{0.65\textwidth}
    \includegraphics[width=0.65\textwidth]{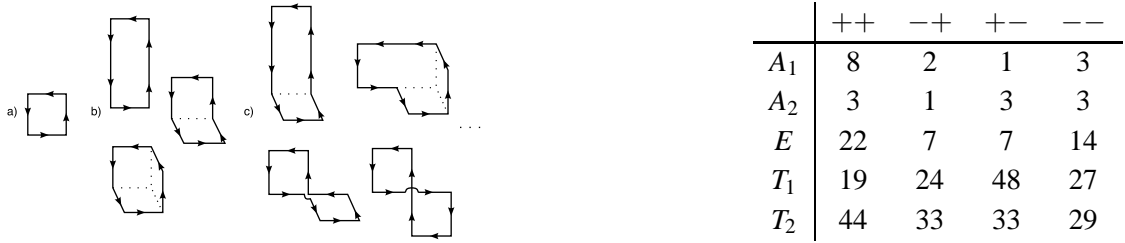}
  \end{minipage}
  \begin{minipage}[.70\textwidth]{.30\textwidth}
    \begin{tabular}[h]{c|cccc}
      & ${++}$ & ${-+}$ & ${+-}$ & ${--}$ \\
      \hline
      $A_1$& 8 & 2 & 1  & 3 \\
      $A_2$& 3  & 1  & 3  & 3\\
      $E$ & 22  & 7  & 7   & 14\\
      $T_1$& 19 & 24 & 48 & 27 \\
      $T_2$& 44 & 33 & 33 & 29 \\
    \end{tabular}
  \end{minipage}
  \caption{(Left) Shapes of basic prototypical paths used to construct
    operators $\mathcal{O}_G(t)$. (Right) Number of single--glueball
    operators included in the variational set for each symmetry
    channel. Each operator is then smeared 4 times.}
  \label{fig:proto-path-glue}
\end{figure}

\begin{figure}[ht!]
 \begin{minipage}[0.15\textwidth]{0.65\textwidth}
    \includegraphics[width=0.65\textwidth]{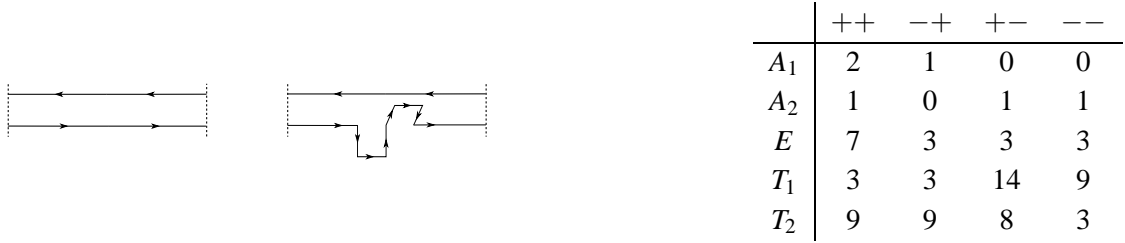}
  \end{minipage}
  \begin{minipage}[.70\textwidth]{.30\textwidth}
    \begin{tabular}[h]{c|cccc}
      & ${++}$ & ${-+}$ & ${+-}$ & ${--}$ \\
      \hline
      $A_1$& 2 & 1 & 0  &  0 \\
      $A_2$& 1 & 0 & 1  &  1 \\
      $E$ & 7 & 3 & 3  &  3 \\
      $T_1$& 3 & 3 & 14 &  9 \\
      $T_2$& 9 & 9 & 8  &  3 \\
    \end{tabular}
  \end{minipage}
  \caption{(Left) Paths used for the construction of operators $\mathcal{O}_S(t)$ coupling with
    torelon states. Periodic Boundary Conditions apply at the edges
    represented by the dashed lines. (Right) Number of $\mathcal{O}_S(t)$
    operators included in the variational set for each symmetry
    channel. Each operator is then smeared 4 times.}
  \label{fig:proto-path-tor}
\end{figure}

On the lattice, the continuum quantum numbers $J^{PC}$ are replaced by
the ones labelling the irreps. of the cubic symmetry group combined
with reflections and charge conjugation, giving a total of $20$
symmetry channels $R^{PC}$. An operator in the channel $R^{PC}$ is
obtained from the
gauge--invariant, vacuum--subtracted operator $\bar{\mathcal{O}}(t)$
by means of
\begin{equation}
  \label{eq:linear-combination}
  \Phi(t) \; = \;  \sum_i c_i \RC_i(\bar{\mathcal{O}}(t))
\quad .
\end{equation}
In the equation above, $\RC_i$ represents a transformation belonging to the full
symmetry group of the system and the coefficients $c_i$ depend
on the channel $R^{PC}$~\cite{Michael:1988jr}.\\
We built three
different classes of operators, one that mainly projects on single--glueball
states, one for two--glueball scattering states and one for torelon excitations.
The single--trace operator that we use to project onto glueball states is simply defined as
\begin{equation}
  \label{eq:glueb-op}
  \mathcal{O}_G(t) \; = \; \frac{1}{N_L^3} \sum_{\vec{x}} \Tr \prod_{l\in\mathcal{C}(\vec{x})}U_l
  \quad .
\end{equation}
In our definition of the variational set we used a wide range of different
closed loops ${\cal C}$, with lengths ranging from $4$ to $8$ lattice
spacing. In Fig.~\ref{fig:proto-path-glue} we summarize the closed loops
used in our simulations and the number of operators built in each
channel. Each of these operators is then smeared and blocked four times. \\
An operator that projects onto scattering states of two glueballs is a double--trace operator. 
Our trial operators for scattering states have the form
\begin{equation}
  \label{eq:double-trace}
  \mathcal{O}_S(t) \; = \; \left( \mathcal{O}_G(t) -
       \VEV{\mathcal{O}_G} \right)^2
  \quad ,
\end{equation}
where we used the definition of the single--trace operator in
Eq.~(\ref{eq:glueb-op}) and the same shapes listed in
Fig.~\ref{fig:proto-path-glue}. The local subtraction of $\VEV{\mathcal{O}_G}$
is crucial in order to obtain the correct two--point function. A more
detailed discussion about the scattering operators can be found in
Ref.~\cite{Lucini:2010nv}.\\
The torelon operators have been created from products of two Polyakov loops
$l_\nu$ winding around opposite directions, in order to obtain an
operator which transforms trivially under the centre of the gauge
group. We defined our operators as
\begin{equation}
  \label{eq:torelon-op}
  \mathcal{O}_T(t) \; = \; \frac{1}{2N_L^2} \sum_{\mu \ne \nu} \sum_x
  l_\nu(x,t)l_\nu^{\dag} (x+\hat{\mu} a,t) 
  \quad ,
\end{equation}
where the sum over $\mu$ runs on the spatial directions orthogonal to
the one of the loops. By choosing different shapes for the combination
$l_\nu(x,t)l_\nu^{\dag} (x+\hat{\mu} a,t)$, we can obtain a fairly
large variational set projecting on torelon states as summarized in
Fig.~\ref{fig:proto-path-tor}.

\section{The results}
\label{sec:results}

\begin{figure}[ht!]
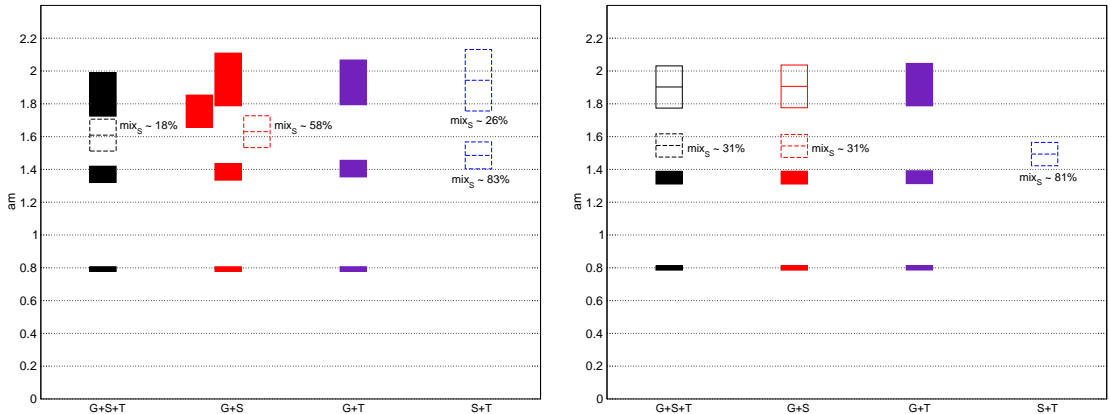

 \centering
  \begin{tabular}{cc}
    \includegraphics[width=0.47\textwidth]{FIGS/A1pp_SU3_cfr_opbasis_200_boxes.eps}&
    \includegraphics[width=0.47\textwidth]{FIGS/A1pp_SU3_L18_cfr_opbasis_200_boxes.eps}\\
  \end{tabular}
 \caption{Variational calculation for
    SU(3) using different sets of operators on a $N_L=12$ lattice (left)
    and a $N_L=18$ lattice (right). The unfilled symbols
    represent masses that cannot be reliably interpreted as pure glueballs.}
  \label{fig:a1pp-3-200}
\end{figure}

The states obtained after the variational procedure can be decomposed
into their projection onto the pure glueball states, the scattering states and onto the
torelons:
\begin{equation}
  \hPh = \sum_i v_i \Phi_i(t) \; \equiv \; \alpha_G \Phi_G + \alpha_S \Phi_S + \alpha_T
  \Phi_T \: ;
  \qquad \mbox{mix}_A = \frac{|\alpha_A|^2}{\sum_i |\alpha_i|^2} \: ;
  \quad A \in \{ G,S,T \} \comma
\end{equation}
where we also defined the relative projection ($\mbox{mix}_A$).
Masses extracted from correlators of $\hPh$ with
$\mbox{mix}_S$,$\mbox{mix}_T \ge 20\%$ can not be
reliably interpreted as pure single--glueball resonances, because
spurious states are expected to propagate between the two sources.\\
After performing the variational calculation, the diagonal elements of
the correlation matrix are fitted with the single-cosh ansatz,
which assumes that only one state dominates the signal. We are often
able to obtain overlaps of the order of $0.95$, which proves the
validity of the original variational ansatz. As a consequence, the fit
generally works very well on the range $ 1 \le t \le 4$~\cite{Lucini:2010nv}.\\
For all gauge groups, there is a high mixing between narrow glueball
trial states and torelon states in the first excitation of the
$E^{++}$ and in the second excitation of the $A_1^{++}$. Other states
with a consistent mixing with the torelons are the $T_2^{+-}$ and the
$T_1^{--}$, the latter mostly for $N=3,4$. Since a calculation
involving scattering states is much more demanding in terms of
computer time, we use the results from the computation involving only
single-particle and torelon operators to target the channels where
mixing with multi-particle states is expected to affect significantly
the results. At large $N$, this is expected to happen for the excited
states that are close to twice the energy of the groundstate. It is
then clear that the channel in which scattering states can potentially
influence the measured spectrum in a relevant way is the $A_1^{++}$,
where we can extract several excitations. We perform calculations on
separate sets of operators in the $A_1$ channel (the full set and the sets obtained
excluding in turn scattering, torelon and single-glueball operators). 
The remarkable property shown by this calculation is that when only
scattering and torelon operators are used the lowest-lying state has a
mass that is much higher (roughly by a factor of two) than the mass of
the groundstate extracted with the full variational basis. Moreover,
the latter appears always when single-particle operators are included
in the calculation. This is an indication that our multi-glueball set
of operators projects only on scattering states, as it should be. The
scattering state seems to be slightly above the first excited
single-glueball excitation at any value of $N$ (see for example the
SU($3$) spectrum of the $A_1^{++}$ channel in
Fig.~\ref{fig:a1pp-3-200}, where two different volumes are
investigated).

\begin{figure}[ht]
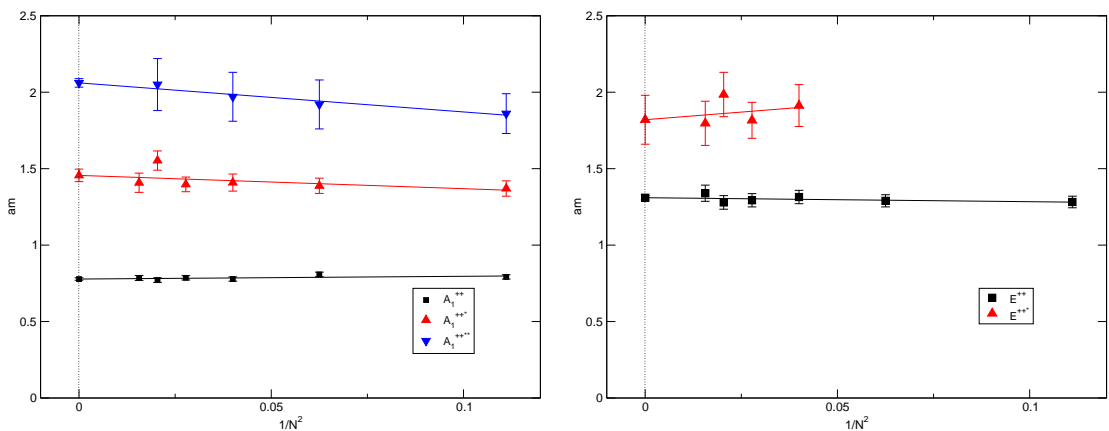

 \centering
  \begin{tabular}{cc}
    \includegraphics[width=0.47\textwidth]{FIGS/A1pp_scatt_fit.eps}&
    \includegraphics[width=0.47\textwidth]{FIGS/Epp_scatt_fit.eps}\\
  \end{tabular}
 \caption{Extrapolation to $N \to \infty$ of the states in the
    $A_1^{++}$ channel (left) and in the $E^{++}$ channel (right).}
  \label{fig:a1pp_ln}
\end{figure}

Having determined the single--glueball spectrum at different values of
$N$ from $3$ to $8$, we extrapolate to the $N=\infty$ limit using the
functional form
\begin{equation}
  \label{eq:lnext}
 a m_G (N)  = a m_G (\infty) + c/N^2 \comma
\end{equation}
dictated by the diagrammatic expansion and already used with success
in Ref.~\cite{Lucini:2004my}. We find that this ansatz
works for all the measured states (including the excitations) for $N
\ge 3$. In general, the central value of $c$ is found to be small
(always of order one or below), as it is expected for a generic
coefficient in a well-behaved expansion. For most of the states we
find only modest corrections to the $N=\infty$ value of the mass: with
a few exceptions, $c$ is compatible with zero and a fit with only the
leading term $a m_G (\infty)$ in Eq.~(\ref{eq:lnext}) gives a result
that is compatible with the fit that includes also the ${\cal
  O}(1/N^2)$ correction~\cite{Lucini:2010nv}. In
Fig.~\ref{fig:a1pp_ln} we show the large--$N$ extrapolation of the
groundstate and the first two excitations of the would--be continuum
scalar $0^{++}$ glueball, together with the groundstate and the first
excited state of the tensor $2^{++}$ glueballs. 

\begin{figure}[ht]
 \centering
  \includegraphics[width=0.70\textwidth]{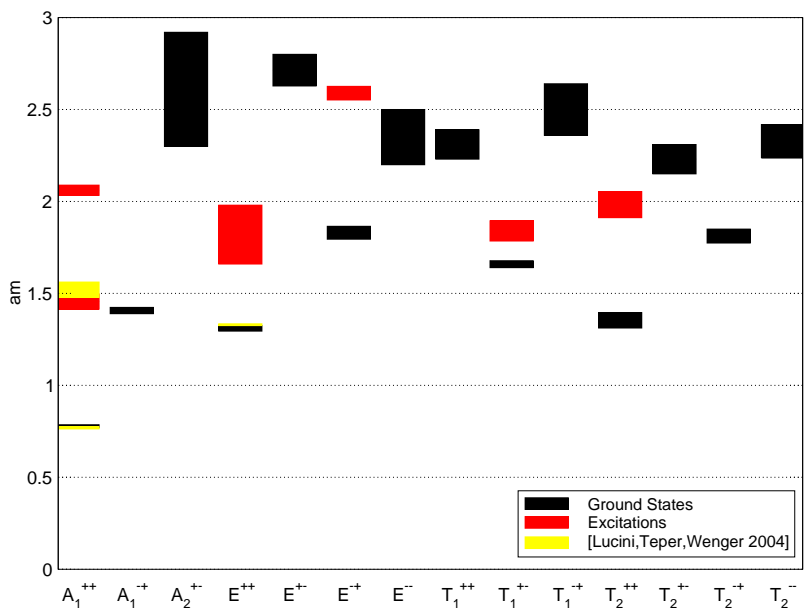}
 \caption{The spectrum at $N=\infty$. The yellow boxes represent the
    large $N$ extrapolation of masses obtained in
    Ref.~\cite{Lucini:2004my}.}
 \label{fig:spectrum_ln}
\end{figure}

The single--glueball spectrum determined in this work is plotted in
Fig.~\ref{fig:spectrum_ln} and it is compared with the known spectrum
at the same lattice spacing taken from
Ref.~\cite{Lucini:2004my}. The latter work achieves a comparable
precision for the  $A_1^{++}$, the $E^{++}$ and the $A_1^{++\star}$, but
in this study we are able to measure seventeen more states. Moreover, the states
present in both studies are compatible.

\section{Conclusions}
\label{sect:conclusions}
In this work, we have studied numerically on the lattice the glueball
spectrum in Yang-Mills SU($N$) gauge theories in the large $N$
limit. Using an automated technique for constructing trial wave
functionals in all possible symmetry channels, we have built a large
variational basis that has enabled us to obtain a large number of
states, including some excitations. Moreover, the inclusion of
functionals that best overlap with scattering and torelon states has
allowed us to unambiguously exclude multi-particle states or
finite-size artefacts from the spectrum of narrow resonances. This is
a significant advance in our understanding of the large $N$ glueball
spectrum from first principles. With little or no modification,
the technique we have presented in this work will also prove helpful
in related problems, like the lattice study of glueballs in QCD and
the study of the low-energy spectrum of confining flux tubes.

\acknowledgments
We thank M. Peardon and M. Teper for discussions on the identification
of scattering states and on the construction of scattering
operators. Discussions with C. McNeile, H. Meyer, C. N\'unez and
A. Patella on various aspects of this work are also gratefully
acknowledged. Numerical simulations have been performed on a 120 core
Beowulf cluster partially funded by the Royal Society and STFC, and on
a 100 core cluster at Wuppertal University. The
work of B.L. is supported by the Royal Society through the University
Research Fellowship scheme and by STFC under contract ST/G000506/1. A.R.
thanks the Deutsche Forschungsgemeinschaft for financial support.
E.R. is supported by a SUPA Prize Studentship. E.R. acknowledges
financial support by the Royal Society in the early
stage of this work.


\end{document}